\UseRawInputEncoding
\documentclass[aps,prc,twocolumn,showkeys,showpacs,groupedaddress,superscriptaddress]{revtex4}
\usepackage{amssymb}
\usepackage{graphicx}% Include figure files
\usepackage{color}

\newcommand{\bel}[1]{\begin{equation}\label{#1}}
\newcommand{\belar}[1]{\begin{eqnarray}\label{#1}}
\def\eps{\epsilon}

\def\pr{\prime}

\begin{document}

\title{The 5-dimensional Langevin approach to fission of atomic nuclei}

\author{F.A. Ivanyuk}
\email{ivanyuk@kinr.kiev.ua}
\affiliation{Institute for Nuclear Research, 03028  Kyiv, Ukraine}
\author{C. Ishizuka}
\email{chikako@nr.titech.ac.jp}
\affiliation{Tokyo Institute of Technology, Tokyo, 152-8550 Japan}
\author{S. Chiba}
\email{chiba.satoshi@nr.titech.ac.jp}
\affiliation{Tokyo Institute of Technology, Tokyo, 152-8550 Japan}
\affiliation{NAT Research Center, NAT Corporation,
3129-45 Hibara Muramatsu, Tokai-mura, Naka-gun,
Ibaraki 319-1112 Japan}

\date{today}

\begin{abstract}
We have generalized the four-dimensional Langevin approach used in our previous works for the description of fission process to the five-dimensional by considering the neck parameter $\eps$ in the two-center shell model shape parametrization as an independent dynamical variable. The calculated results for the mass distribution of fission fragments are in better agreement with the available experimental data. In particular, the transition from the mass-symmetric to mass-asymmetric fission via the triple-humped distribution in fission of thorium isotopes is well reproduced.
\end{abstract}


\maketitle

%%%%%%%%%%%%%%%%%%%%%%%%%%%%%%%%%%%%%%%%%%%%%%%%%%%%%%%%%%%%%%%%%%%%%%%%%%%%%%%%%%%%%%%%%%%%%%%%%%%%%%%
\section{Introduction}
\label{intro}
%%%%%%%%%%%%%%%%%%%%%%%%%%%%%%%%%%%%%%%%%%%%%%%%%%%%%%%%%%%%%%%%%%%%%%%%%%%%%%%%%%%%%%%%%%%%%%%%%%%%%%%
The approach based on Langevin equations \cite{langevin} has been successfully applied in various branches of theoretical physics and chemistry for many years. In nuclear physics, this approach is  used for the description of fission or fusion  processes at excitations above the fission barrier  \cite{nix76,wada93,froebrich,pomorski1996,karpov,asano,adeev2005,aritomo,mazurek,our17,scirep,pasha,kosenko}.
In these works the Langevin equation was solved in 1-4 dimensions
with macroscopic \cite{werwhe,wall1,runswiat} or microscopic \cite{hofrep,hofbook,ivahof} transport coefficients.

The approach describes quite well the mass distributions and kinetic energies of fission fragments, the multiplicities of emitted neutrons, and other observables of fission or fusion  processes.

Five-dimensional calculations  were published so far only by Sierk \cite{sierk} using the three-quadratic-surface shape parametrization. In that work the potential energy was calculated within the macroscopic-microscopic model, the inertia in Werner-Wheeler approximation, and the friction by "surface-plus-window" formula with the reduction factor $k_s$=0.27. Because the numerical calculations were very time consuming,
the starting point was chosen slightly outside of the outermost saddle point (second or third barrier) and the initial velocity was picked up randomly from the Kramers velocity distribution for the local temperature at the saddle point deformation. Only velocities directed towards scission were taken into account.
The scission in this model was defined as occurring when the minimum neck radius falls to a specific value $r_{neck}^{(crit)}$=1 fm.

Within this model a number of observables, such as the mean mass asymmetry seen in fission, the approximate width of the mass yields of the heavy and light peaks, the approximate average fragment kinetic energy and width for fission of actinide nuclei both spontaneous and induced by neutrons of energies of up to the threshold for second-chance fission were accurately reproduced.

The examination of fission of thorium and radium isotopes, which exhibits a transition to yield  distributions where symmetric fission  is of a  magnitude comparable to asymmetric fission, was mentioned as one of the most interesting explorations of the model.

Recently the first results of 5D Langevin calculations with deformed Woods-Saxon potential and Cassini shape parametrization were reported by T. Wada at the Kazimierz workshop \footnote{29th Nuclear Physics Workshop, 26-30 September 2023, Kazimierz Dolny, Poland}.

In present work we have generalized our four-dimen\-sio\-nal Langevin approach to the five-dimensional in order to explain the transition from the mass-symmetric to mass-asymmetric fission of thorium isotopes.

\begin{figure}[ht]
\centering
\includegraphics[width=0.75\columnwidth]{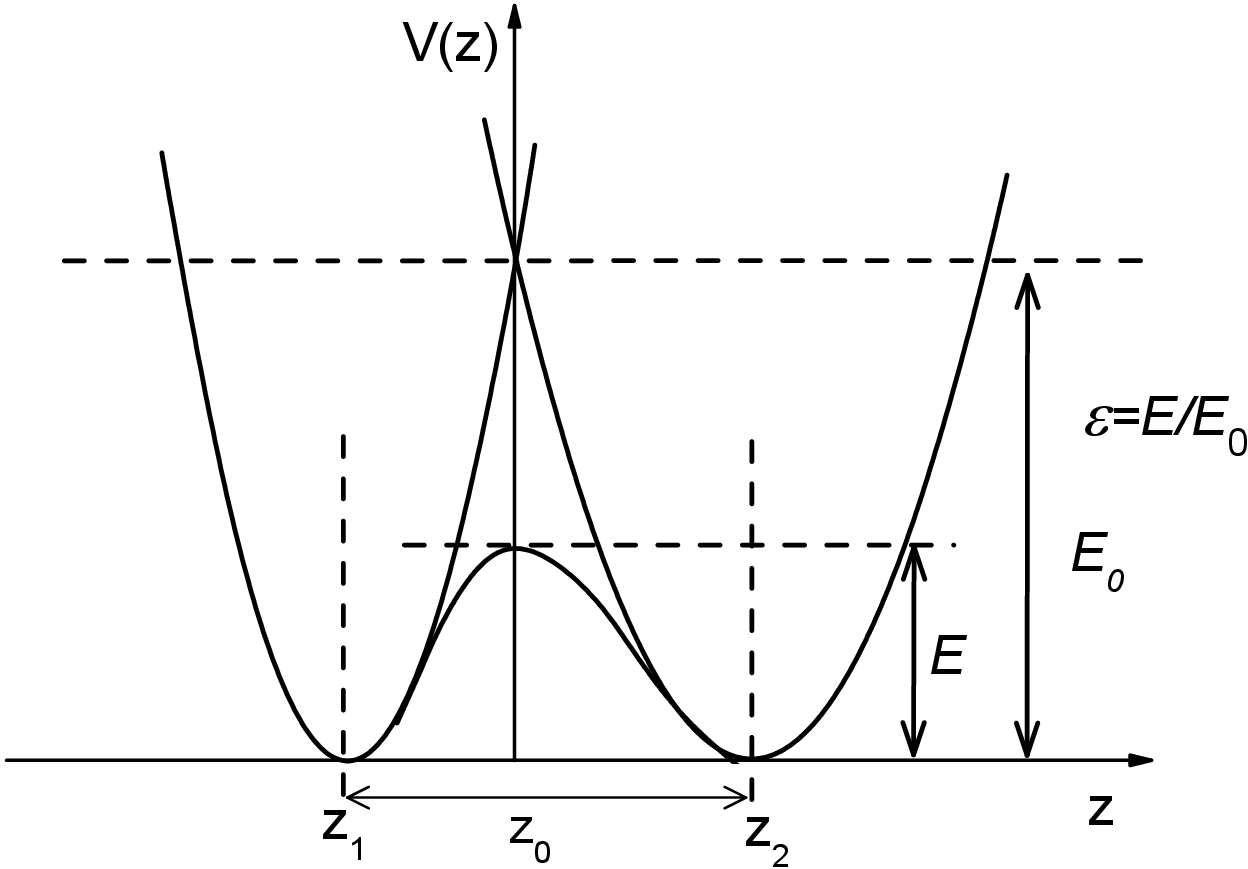}
\caption{The $z$-dependent part of the TCSM mean-field potential $V(z)\equiv V(\rho,z)_{\rho=\rho(z)}$. The neck parameter $\eps$ is defined as the ratio of the potential $V(z)$ at $z=0$ to the value $V_0$ of oscillator potential at $z=0,  \eps\equiv V(z=0)/V_0$.}
\label{definit}
\end{figure}

In the Langevin approach one solves the set of differential equations for the time evolution of collective variables $q_{\mu}$
describing the shape of the nuclear surface. In our works we used for the shape
parametrization that of the two-center shell model (TCSM) \cite{tcsm}, see Fig.\,\ref{definit}. In this model the shape of the axially symmetric surface is characterized by 5 deformation parameters $q_{\mu} =z_0/R_0, \delta_1, \delta_2, \alpha $ and $\epsilon$.
Here z$_0/R_0$ refers to the distance
between the centers of left and right oscillator potentials, $R_{0}$ being the radius of spherical nucleus. The parameters $\delta_1$ and $\delta_2$ describe the
deformation of the right and left parts of the nucleus. The fourth parameter
$\alpha $ is the mass asymmetry and the fifth parameter of TCSM shape
parametrization $\epsilon$ regulates the neck radius.

In our older Langevin calculations \cite{aritomo,our17}
we used a three-dimensional shape parametrization
in which $\delta_1$ and $\delta_2$ were assumed to be equal, $\delta_1=\delta_2$, and the neck parameter was kept constant, $\eps=0.35$.
The generalization from three to four dimensions was carried out in \cite{four}. In that work the deformation parameters $\delta_1$ and $\delta_2$ were considered as an independent dynamical variables. Thus, the shapes with very different deformations of the left and right parts were included into the consideration. In particular, one part could be nearly spherical while the other is very elongated.
This made it possible to reproduce \cite{fermis} the rapid change of mass distributions of fission fragments from mass-asymmetric (in $^{256}$Fm) to mass-symmetric (in $^{258}$Fm) \cite{fermis},  the saw-tooth structure of neutron multiplicity \cite{multy}, and the decrease of the total excitation energy of fission fragments with the increasing of the excitation energy \cite{shimada}.
%%%%%%%%%%%%%%%%%%%%%%%%%%%%%%%%%%%%%%%%%%%%%%%%%%%%%%%%%%%%%%%%%
%%%%%%%%%%%%%%%%%%%%%%%%%%%%%%%%%%%%%%%%%%%%%%%%%%%%%%%%%%%%%%%%%%%%%%%%%%%%%%%%%%%%%

At the same time, there are some experimental results on nuclear fission that are not explained theoretically up to now.
In \cite{khs} the secondary-beam facility of GSI Darmstadt was used to study the fission properties of 70 short-lived radioactive nuclei.  Relativistic secondary projectiles were produced by fragmentation of a 1 A GeV
$^{238}$U primary beam and identified in nuclear charge and mass number. These reaction products  were excited by electromagnetic interactions, and fission from excitation energies around 11 MeV was induced. The elemental yields and the total kinetic energies for a series of neutron-deficient pre-actinides and actinides from $^{205}$At to $^{234}$U were determined. The elemental yields after electromagnetic-induced fission, cover the transition from symmetric fission at $^{221}$Ac to asymmetric fission at $^{234}$U. The longest isotopic sequence, from $^{217}$Th to $^{229}$Th, was measured for thorium isotopes.
In the transitional region, around $^{227}$Th, triple-humped distributions appear, demonstrating comparable weights for symmetric and asymmetric fission.

The qualitative transition from the mass-symmetric to mass-asymmetric mass distributions for thorium isotopes was shown in the first paper on the random walk model \cite{random} assuming the overdamped character of nuclear collective motion.
Later (after adding the energy dependence
which has some effect at E*=11 MeV) much better agreement of the calculated fission-fragment charge yields with GSI data \cite{khs} was demonstrated in \cite{randrup2013}.
The transition from mass-symmetric to mass-asymmetric in sequence of thorium isotopes was also found recently by the Polish-Chinese collaboration \cite{pomo2021} within the Born-Oppenheimer approximation.

A more recent experimental study of nuclear fission along the thorium isotopic sequence was carried out in \cite{chatil1}.
In that article, the new data on fission along the thorium isotopic sequence were reported, using the same reaction mechanism as in \cite{khs}, but with the new R3B/SOFIA experimental setup which was conceived to identify the mass and the nuclear charge of the fissioning nuclei and both fission fragments.
%%%%%%%%%%%%%%%%%%%%%%%%%%%%%%%%%%%%%%%%%%%%%%%%%%%%%%%%%%%%%%%%%
\begin{figure}[ht]
\centering
\includegraphics[width=0.95\columnwidth]{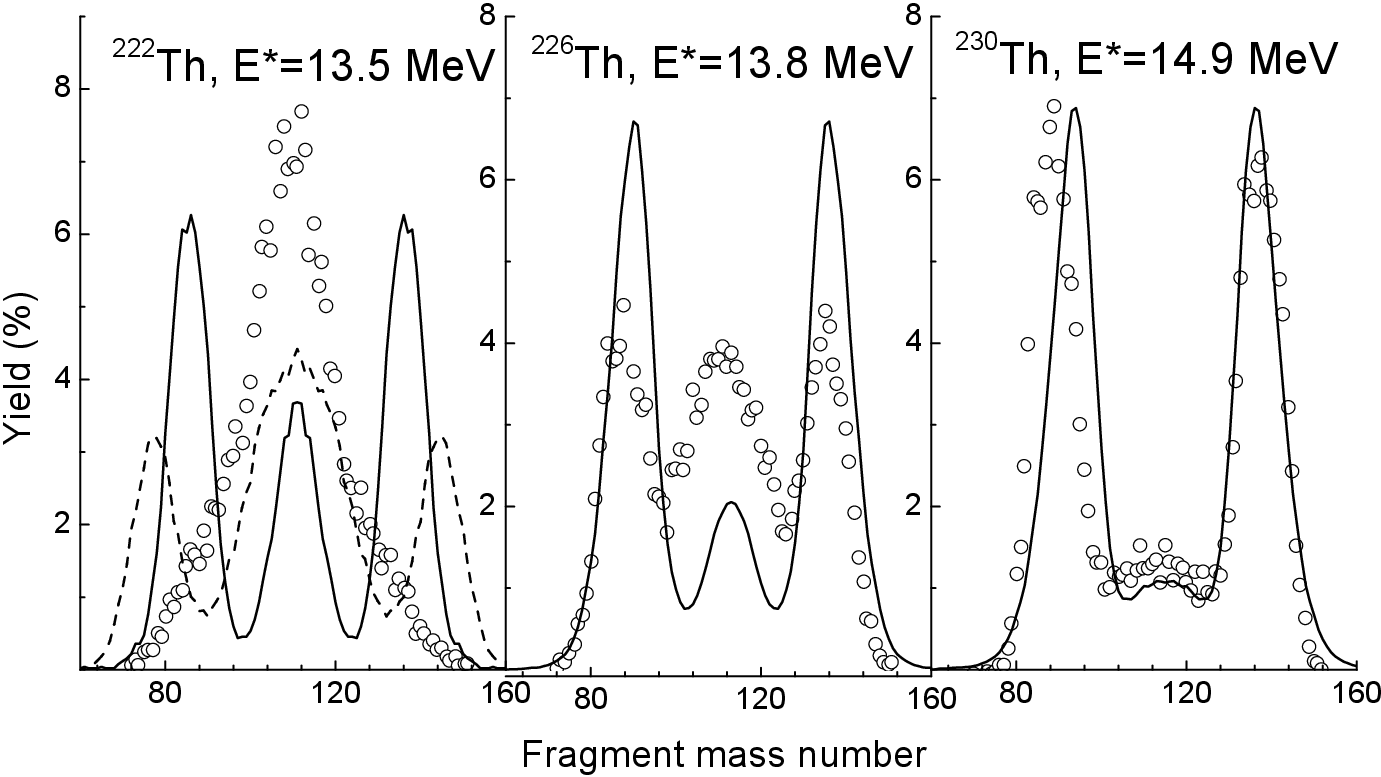}
\caption{The fission fragment mass distributions for few thorium isotopes. The experimental data from \protect\cite{chatil1} are marked by circles, solid lines show the 4D Langevin calculations with $\eps$=0.35, and the dashed line shows the 4D Langevin calculation for $^{222}$Th  with $\eps$=0.15.}
\label{4D_Chatil}
\end{figure}
%%%%%%%%%%%%%%%%%%%%%%%%%%%%%%%%%%%%%%%%%%%%%%%%%%%%%%%%%%%%%%%%%%%%%%%%%%%%%%%%%%%%%

The experimental data from \cite{chatil1} for few thorium isotopes are shown by circles in Fig.\,\ref{4D_Chatil}. Solid lines show the 4D Langevin calculations with $\eps$=0.35. One can see that for $^{230}$Th the agreement of experimental and theoretical results is quite reasonable. For lighter isotopes the experimental mass symmetric peak becomes stronger and stronger, while in calculations the mass asymmetric peak remains dominant.

As one can see from Fig.\,\ref{4D_Chatil}, the mass distribution calculated with the four-dimensional approach is mainly
 mass asymmetric. The mass symmetric peak is too small. All attempts to make it larger varying the starting point or the scission conditions were not successful. The only noticeable effect on the height of central peak arose from a variation
of the neck parameter $\eps$. With smaller $\eps$ the mass symmetric peak became somewhat larger, see the dash curve in the left part of Fig.\,\ref{4D_Chatil}.
That was a clear indication that the parameter $\eps$ of TCSM shape parametrization should not be fixed, but should be considered as an independent dynamical variable, like the other four deformation parameters.

The results of the five-dimensional Langevin calculations for the sequence of thorium isotopes are presented below in this publication.

In Section II we present the main relations of the Langevin approach to the description of nuclear fission process.  In Section III we explain how the collective potential energy is calculated. Section IV contains the description of the used mass and friction tensors. The results of numerical 5D calculation are given in Section V. Section VI contains short summary.

%%%%%%%%%%%%%%%%%%%%%%%%%%%%%%%%%%%%%%%%%%%%%%%%%%%%%%%%%%%%%%%%%%%%%%%%%%%%%%%%%%%%%%%%%%%%%%%%%%%%%%%
\section{The Langevin approach}
\label{model}
%%%%%%%%%%%%%%%%%%%%%%%%%%%%%%%%%%%%%%%%%%%%%%%%%%%%%%%%%%%%%%%%%%%%%%%%%%%%%%%%%%%%%%%%%%%%%%%%%%%%%
The first-order differential equations (Langevin equations) for the time
dependence of the collective variables $q_{\mu }$ and the conjugate momenta
$p_{\mu }$ are:

\belar{lange}
\frac{dq_\mu}{dt}&=&\left(m^{-1} \right)_{\mu \nu} p_\nu , \\
\frac{dp_\mu}{dt}&=&-\frac{\partial F(q,T)}{\partial q_\mu} - \frac{1}{2}\frac{\partial (m^{-1})_{\nu \sigma} }{\partial q_\mu} p_\nu p_\sigma
 -\gamma_{\mu \nu}(m^{-1})_{\nu \sigma} p_\sigma \nonumber\\
 &+&g_{\mu \nu} R_\nu (t),
\end{eqnarray}
where the sums over the repeated indices are assumed. In Eq.\,(2)  the $F(q, T)$ is the
temperature dependent free energy of the system, $\gamma _{\mu \nu }$
and ($m^{-1})_{\mu \nu }$ are the friction and inverse of mass tensors, and $g_{\mu \nu}R_{\nu}$(t) is the random force.

The free energy $F(q, T)$ is calculated as the sum of the macroscopic (folded Yukawa) energy
and the temperature dependent shell correction $\delta F(q, T)$.
The shell corrections are calculated from the single-particle energies in the deformed
Woods-Saxon potential
\cite{pash1,pash2} fitted to the mentioned above TCSM shape
parameterizations.

The collective inertia tensor $m_{\mu \nu}$ is calculated within the
Werner-Wheeler approximation \cite{werwhe}
and for the friction tensor $\gamma_{\mu \nu }$ we
used the wall-and-window formula  \cite{wall1,runswiat}.

The random force g$_{\mu \nu}R_{\nu }$(t) is the product of the normally distributed white noise
$R_{\nu}$(t), $\langle R_{\mu}(t)R_{\nu}(t^{\pr})\rangle=2 \delta_{\mu\nu}\delta (t-t^{\pr})$, and the temperature-dependent strength factors
$g_{\mu \nu}$. The factors g$_{\mu \nu }$ are related to the temperature
and friction tensor via the modified Einstein relation,
\begin{equation}\label{teff}
g_{\mu \sigma } g_{\sigma \nu } =T^\ast \gamma _{\mu \nu }
\,,\,\,\,\,\,\,\,\,\,\,\,\,\,\,\,\,\,\,\,\,\,\,T^\ast
=\frac{\hbar \varpi}{2}\coth \frac{\hbar \varpi}{2T}\,\,,\,\,\,\,
\end{equation}
where $T^{\ast }$ is the effective temperature \cite{pomhof,hofkid}. The parameter $\varpi$
 is the local frequency of collective motion \cite{hofkid}. The minimum of
$T^{\ast }$ is given by $\hbar \varpi/2$. At large excitations $T^{\ast }$ is close to $T$.

In \cite{pomhof,hofkid} the
one-dimensional collective motion was considered and the "frequency
of collective motion" was simply the frequency of harmonic
vibrations around a fixed point in the deformation space. In the considered here model
 the collective space is five-dimensional, so
there are five collective frequencies at each deformation point.
Besides, these frequencies depend on the point in the
collective space.

Unfortunately, with our present computational
facility, we can not take into account the dependence of $\hbar \varpi$
on deformation and the type of collective degree of freedom. So,
we use the simplified approximation, namely, the constant value for $\hbar \varpi$, the same for all degrees of freedom.

To check how sensitive are the calculated mass distributions to the
choice of $\hbar \varpi$, we have carried out the calculations with
three values of $\hbar \varpi$: $\varpi$=0 ($T^{\ast}=T$), $\hbar \varpi$=2 and $\hbar \varpi$=4
Mev, please, see the red, black and blue curves in Fig.\,\ref{5D_Chatil}. It
turned out the mass distributions of $^{222}$Th, $^{226}$Th and $^{230}$Th
calculated with $\varpi$=0, $\hbar \varpi$=2 and 4 Mev are surprisingly close to
each other. So, all other calculations in the present work were carried out with
$\hbar \varpi$=2 Mev. The same value for $\hbar \varpi$ was used also in our previous publications.

%In principle, $\varpi$ should depend on deformation.
%With larger $\varpi$ the effective temperature and the friction force become larger and the width of the peaks of mass %distributions gets broader (at small excitations). Unfortunately, the account of the deformation dependence of $\varpi$ makes the computations too time consuming. In our previous  works we used for $\varpi$ the deformation-independent value $\hbar\varpi$=2 MeV. The same value was used in the present calculations. As one can see from Fig.\,\ref{5D_Chatil} below, the width of calculated and experimental distributions are rather close to each other.

A discussion of the
zero-point energy in different models can be found in the seminal paper
by Hill and Wheeler \cite{hillw}.
The approximate value seems to be between 0.5 to 2.23 MeV. Note, that the zero-point energy is equal to $\hbar\varpi$/2.

The temperature $T$ in Eq.(\ref{teff}) is related to the initial energy $E^{\ast}_{(in)}$
and the local excitation energy $E^{\ast}$ by,
\begin{equation}\label{Exx}
E^\ast =E_{gs} +E^\ast _{(in)} -\frac{1}{2}m^{-1}_{\mu \nu } p_\mu p_\nu
-V_{pot} (q,T=0)=aT^2,
\end{equation}
where $V_{pot}$ is the potential energy and $a$ is the level density parameter.
More details are given in our earlier publications, see \cite{14,24,four,26}.
The $E^{\ast}_{(in)}$ in Eq.\,\ref{Exx} is the energy of initial excitation of nuclei by the Coulomb excitation,
see \cite{chatil1}. In case of neutron induced fission
$E^{\ast}_{(in)}$  would be the sum of the neutron kinetic and separation energies.

Usually, the initial  values of momenta $p_{\mu}$ are set to zero, and calculations are
started from the ground state deformation.

The calculations are continued until the trajectories reach the "scission point". In our older publications the "scission point" was defined as the point in deformation space where the neck radius becomes zero. However, the zero critical neck radius is not well justified. It was shown in the so called optimal shapes model \cite{strut1963,iva2009,pomiva2} that the nuclear liquid drop loses stability with respect to elongation at an almost constant elongation $z_0\approx$ 2.3 $R_0$ and rather thick neck, $r_{neck}^{crit}$, see right part of Fig. 1 in \cite{iva2009}.
%see Fig.\,\ref{optimal}.
Thus, in all our calculations below we used the finite value of the critical neck, $r_{neck}^{crit}$ = 1 fm.
%%%%%%%%%%%%%%%%%%%%%%%%%%%%%%%%%%%%%%%%%%%%%%%%%%%%%%%%%%%%%%%%%
%\begin{figure}[ht]
%\centering
%\includegraphics[width=0.75\columnwidth]{fig1b.eps}
%left and right parts of nucleus, and the shapes at the scission point. The $x_{LD}$ is the fissility parameter, the ratio of Coulomb and surface energy for spherical shape, $x_{LD}\equiv E_{Coul}^{(sph}/E_{surf}^{(sph)}$.}
%\label{optimal}
%\end{figure}
%%%%%%%%%%%%%%%%%%%%%%%%%%%%%%%%%%%%%%%%%%%%%%%%%%%%%%%%%%%%%%%%%%%%%%%%%%%%%%%%%%%%%
%%%%%%%%%%%%
%%%%%%%%%%%%%%%%%%%%%%%%%%%%%%%%%%%%%%%%%%%%%%%%%%%%%%%%%%%%%%%%%%%%%%%%%%%%%%%%%%%%%%%%%%%%%%%%%%%%%%%
\section{The 5D shape parametrization}
\label{5Dparams}
%%%%%%%%%%%%%%%%%%%%%%%%%%%%%%%%%%%%%%%%%%%%%%%%%%%%%%%%%%%%%%%%%
In present  work we use the two center shell model suggested by Maruhn and Greiner \cite{tcsm} and the code developed by Suekane, Iwamoto, Yamaji and Harada \cite{suek74,iwam76,sato79} and extended by one of the authors (Ivanyuk, \cite{four}).

The TCSM single-particle Hamiltonian $H$ includes the mean-field potential $V(\rho,z)$ and the angular momentum dependent part.
In cylindrical coordinates $\{\rho, z\}$ it is written as
\begin{equation}\label{hamil}
H=\frac{\vec p~^2}{2m} + V(\rho, z)-\kappa_i[2(\vec l_i \cdot \vec s)+\mu_i (\vec l_i^2-<\vec l_i~^2>)]\hbar\omega_0,
\end{equation}
see \cite{suek74,iwam76,sato79}. Here $i=1$ for $z\leq 0$ and $i=2$ for $z\geq 0$, $\kappa_i$ and $\mu_i$ are the usual parameters of Nilsson model \cite{nilsson}.

The potential $V(\rho,z)$ in TCSM consists of the two oscillator potentials with centers at $z_1$ and $z_2$, smoothly joined between $z_1$ and $z_2$ by a fourth order polynomial in $z$, see Fig.\,\ref{definit}. The deformation parameters $\delta_1$ and $\delta_2$ define the curvature of the parts outside of $z_1$ or $z_2$. In the three-dimensional parametrization $\delta_1$ and $\delta_2$ are the same.

The advantage of the TCSM shape parametrization is that even in the case of three deformation parameters ($\delta_1=\delta_2$) it supplies a very reasonable shape parametrization for large deformations, including the shape of separated fragments. The three-dimensional Langevin calculations are not very time-consuming. They are successfully used up to now for the description of the fission process \cite{yoshi2022,yoshi2023,lile2022,huang2022}. However, the 3D shape parametrization can not describe the shapes where one part of the nucleus is close to a sphere and the other is very elongated. This drawback was corrected in \cite{four} by the generalization of 3D to 4D Langevin approach.

%The TCSM shape parametrization meets some difficulty at small deformation (close to spherical shape) and large mass asymmetry. To overcome this drawback a modification of the standard TCSM shape parametrization was suggested in \cite{24}. The modified TCSM shapes are quite smooth and the ground state energy is defined rather accurately. One should keep also in mind that the value of the ground state energy appears only in the definition of the initial excitation energy and has a very small effect on the mass distribution of the fission fragments.

\begin{figure}[ht]
\centering
\includegraphics[width=0.85\columnwidth]{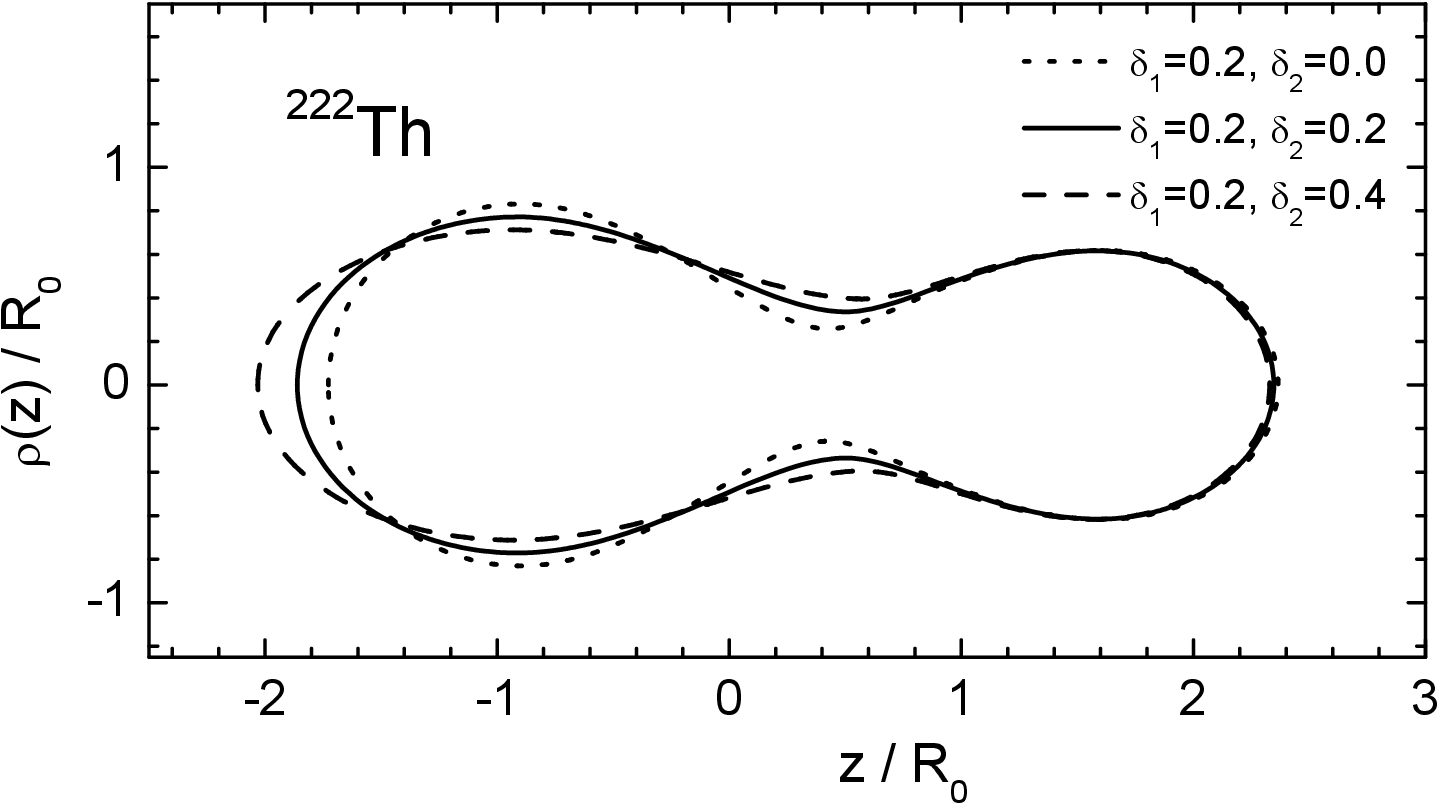}
\caption{The effect of variation of parameter $\delta_2$ on 4D TCSM shapes, $z_0/R_0$=2.5, $\alpha$=0.2, $\eps$=0.35.}
\label{4Dshapes}
\end{figure}

The variation of $\delta_1$ at fixed $\delta_2$ formally change only the outer part of the potential. But, due to the continuity conditions at $z_1$ and $z_2$, requirements of fixed volume and mass asymmetry, the whole shape, even in the neck region, is modified by variation of $\delta_1$ or $\delta_2$, see Fig.\,\ref{4Dshapes}. So, $\delta_1$ and $\delta_2$ can be considered as the parameters of deformation of the entire left and right parts of the shape.
%%%%%%%%%%%%%%%%%%%%%%%%%%%%%%%%%%%%%%%%%%%%%%%%%%%%%%%%%%%%%%%%%%%%%%%%%%%%%%%%%%%%%
%%%%%%%%%%%%%%%%%%%%%%%%%%%%%%%%%%%%%%%%%%%%%%%%%%%%%%%%%%%%%%%%%%%%%%%%%%%%%%%%%%%%%

\begin{figure}[ht]
\centering
\includegraphics[width=0.85\columnwidth]{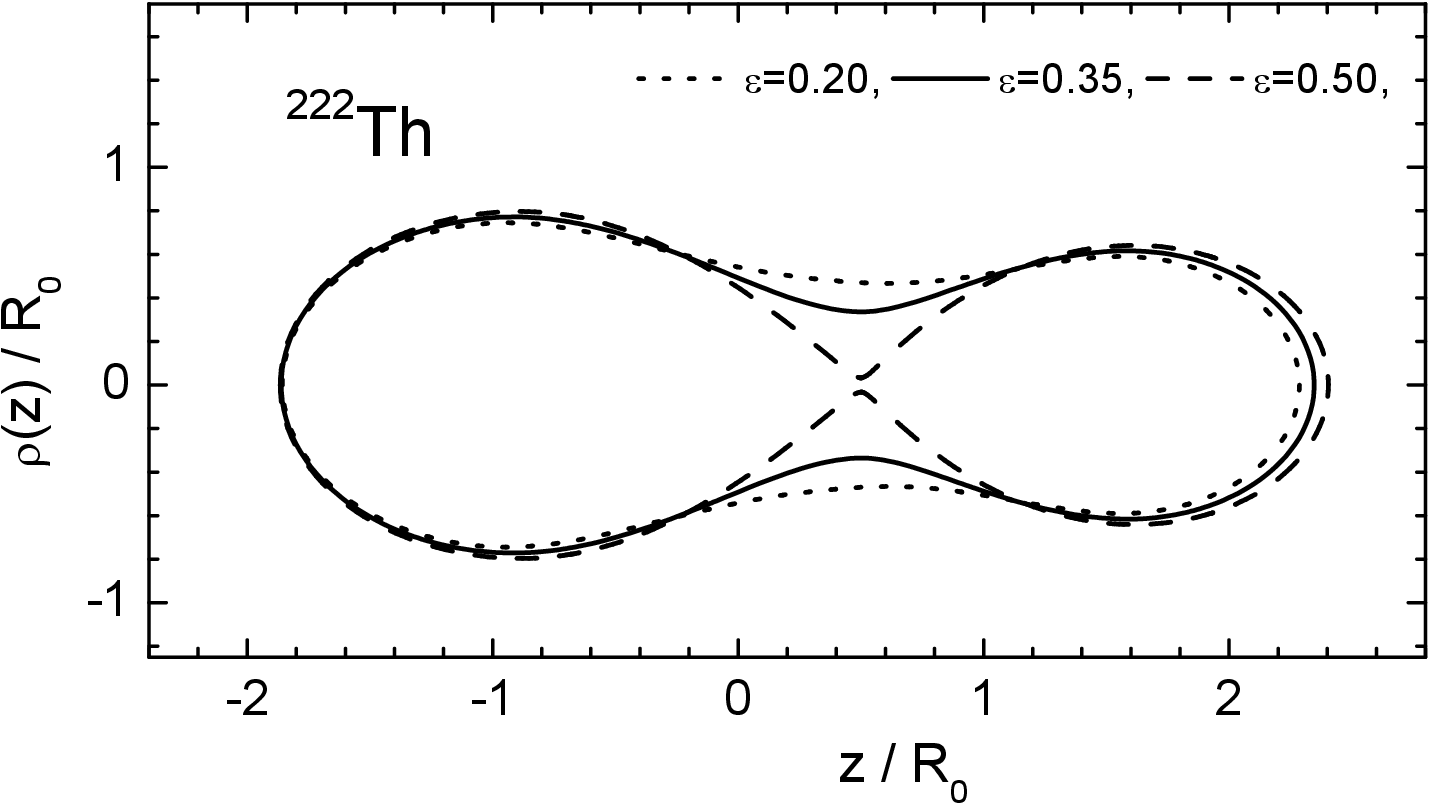}
\caption{The effect of variation of parameter $\eps$ on 5D TCSM shapes, $z_0/R_0$=2.5, $\delta_1=\delta_2=\alpha$=0.2.}
\label{5Dshapes}
\end{figure}

The effect of the variation of the fifth parameter, $\eps$, on the nuclear shape is somewhat different. Formally, parameter $\eps$ controls the neck radius, see Fig.\,\ref{5Dshapes}.

%%%%%%%%%%%%%%%%%%%%%%%%%%%%%%%%%%%%%%%%%%%%%%%%%%%%%%%%%%%%%%%%%
\begin{figure}[ht]
\centering
\includegraphics[width=0.85\columnwidth]{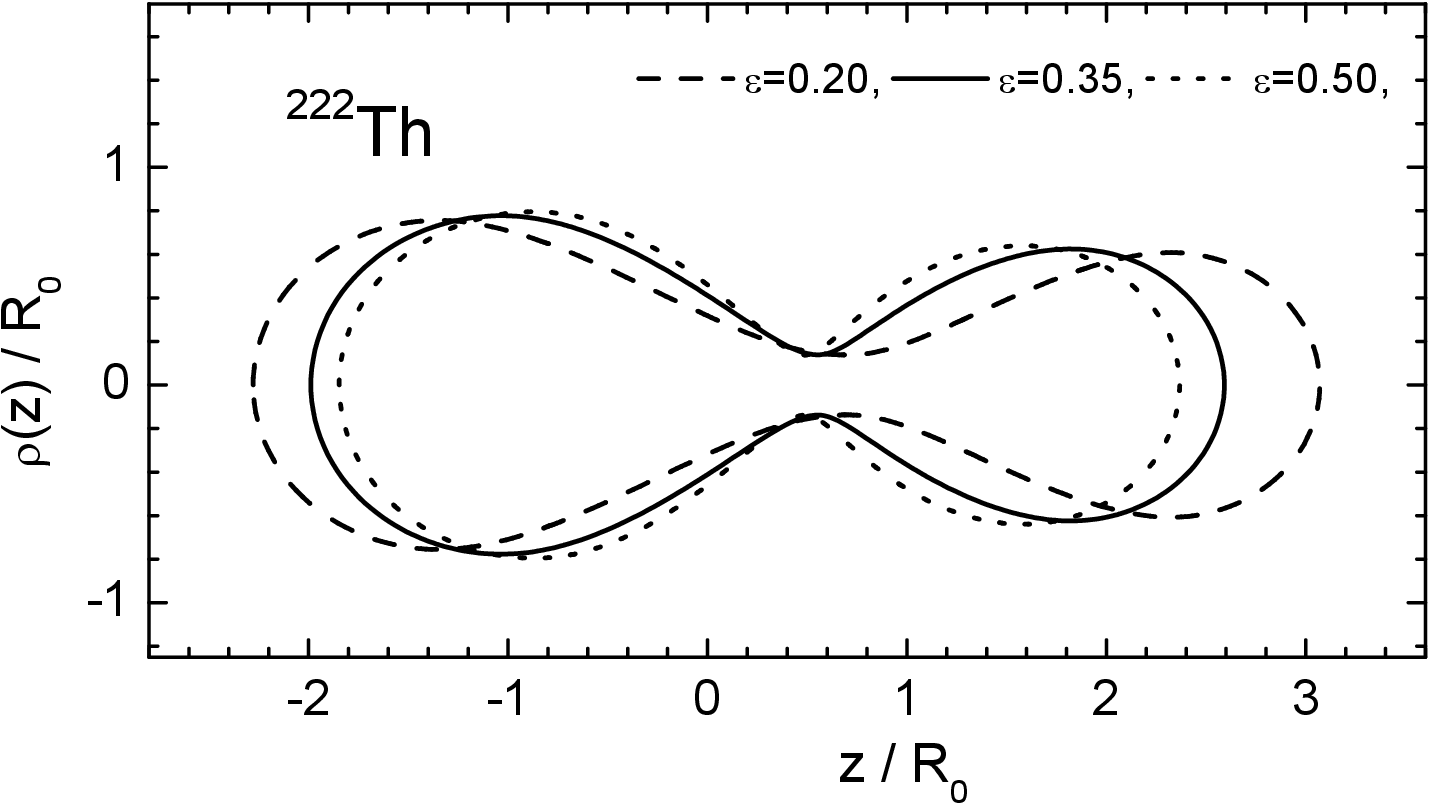}
\caption{The 5D TCSM shapes for $\delta_1=\delta_2=\alpha$=0.2 at the scission point for few values of parameter $\eps$}
\label{sci_shapes}
\end{figure}

%%%%%%%%%%%%%%%%%%%%%%%%%%%%%%%%%%%%%%%%%%%%%%%%%%%%%%%%%%%%%%%%%%%%%%%%%%%%%%%%%%%%%
However, since the yields and other quantities are calculated at the scission point with a fixed neck radius, the variation of $\eps$ leads to the incorporation of very elongated or very compact shapes, see Fig.\,\ref{sci_shapes}.
%%%%%%%%%%%%%%%%%%%%%%%%%%%%%%%%%%%%%%%%%%%%%%%%%%%%%%%%%%%%%%%%%
%%%%%%%%%%%%%%%%%%%%%%%%%%%%%%%%%%%%%%%%%%%%%%%%%%%%%%%%%%%%%%%%%%%%%%%%%%%%%%%%%%%%%
Exactly such shapes contribute to the super-long or super-short fission modes. So, the inclusion of $\eps$ as an additional dynamical variables makes it possible to describe more accurately the super-long or super-short fission modes.
%%%%%%%%%%%%%%%%%%%%%%%%%%%%%%%%%%%%%%%%%%%%%%%%%%%%%%%%%%%%%%%%%%%%%%%%%%%%%%%%%%%%%%%%%%%%%%%%%%%%%%%
\section{The potential energy}
\label{poten}
%%%%%%%%%%%%%%%%%%%%%%%%%%%%%%%%%%%%%%%%%%%%%%%%%%%%%%%%%%%%%%%%%
The parameters $\kappa$ and $\mu$ of Nilsson model in (\ref{hamil}) were fitted separately for each oscillator shell \cite{nilsson} in order to reproduce the known single-particle energies at the ground state of nuclei. It is evident that $\kappa$ and $\mu$ may depend on deformation. Away from the ground state, at the barrier and beyond, $\kappa$ and $\mu$ may differ from these at the ground state. Thus, the use of TCSM with the Nilsson potential is not well justified for large deformations.  Consequently, we have modified the mean-field potential of TCSM. Instead of the Nilsson type of potential we use a
more realistic Woods-Saxon (TCWS) potential \cite{four}. For this, we keep from TCSM only the deformation dependent profile function $\rho(z)$ (the axially symmetric nuclear shape is obtained by the rotation of $\rho(z)$ around the $z$ axis). Then the profile function $\rho(z)$ is expanded in a series of Cassini ovaloids.
In total, 20 terms in the expansion are taken into account, which makes the reproduction of the TCSM shapes very accurate, see Fig.\,3 of \cite{four}.

Considering the coefficients  $\alpha_n$ of the expansion as the deformation parameters of Woods-Saxon potential, the single-particle wave functions and energies are calculated with the code of Pashkevich \cite{pash1,pash2}.

The free energy of nucleus $F(q)$ is calculated within the macroscopic-microscopic model,
\begin{equation}\label{free}
F(q)=E_{LDM}(q)+\delta F(q,T)
\end{equation}

The shell correction $\delta F(q,T)$ is calculated by Strutinsky's prescription \cite{struti,brdapa} from the energies of single-particle states in the deformed Woods-Saxon potential fitted to the TCSM shapes.
At zero temperature the shell correction to the free energy $\delta F(q,T=0)$ coincides with the shell correction to the collective potential
energy $\delta E(q)$,
%The shell correction $\delta E(q,T)$  contains the  contributions from the shell effects in total single-particle energy and in the pairing energy.
\begin{equation}
\label{deltae}
\delta E(q) =\sum_{n,p} \left( \delta E_{shell}^{(n,p)}(q) + \delta E_{pair}^{(n,p)} (q) \right) .
\end{equation}

The damping of $\delta F(q, T)$ with the excitation energy was calculated by the method developed in \cite{shcot}.

The macroscopic part of energy $E_{LDM}(q)$ is calculated within the folded Yukawa model \cite{suek74,iwam76}.

\begin{figure}[ht]
\centering
\includegraphics[width=0.95\columnwidth]{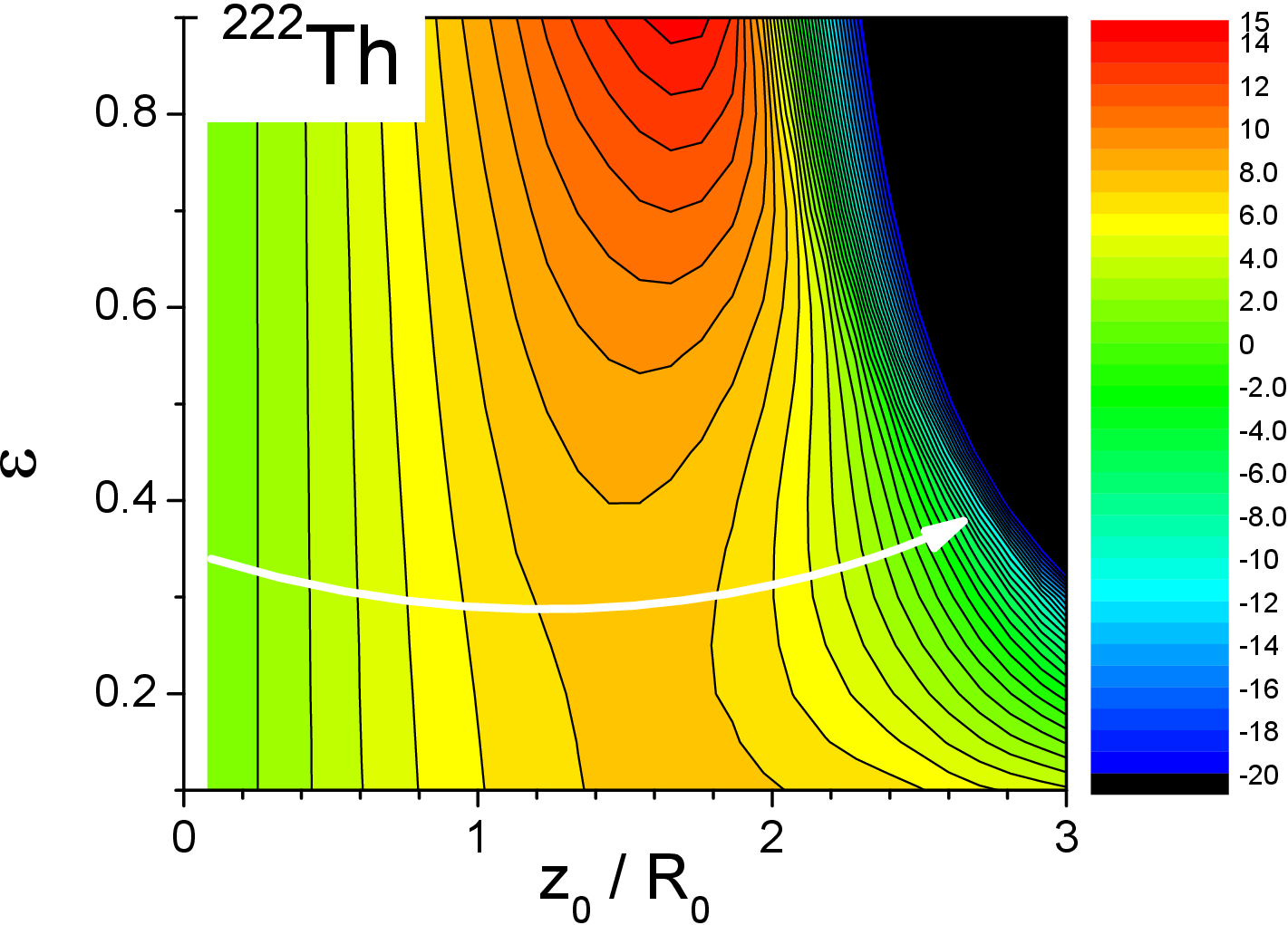}
\caption{The dependence of macroscopic deformation energy of $^{222}$Th on the elongation $z_0/R_0$ and neck parameter $\eps$ for fixed $\delta_1=\delta_2=\alpha$=0.2. }
\label{edef15}
\end{figure}

An example of the macroscopic deformation energy $E_{LDM}(q)$ is shown in Fig.\,\ref{edef15}. As one can see, the potential energy is rather flat in $\eps$ direction. Consequently, the dynamical trajectories fill all the available space in $\eps$.
%%%%%%%%%%%%%%%%%%%%%%%%%%%%%%%%%%%%%%%%%%%%%%%%%%%%%%%%%%%%%%%%%%%%%%%%%%%%%%%%%%%%%%%%%%%%%%%%%%%%%%%%%%%%%%%%%%%%%
%%%%%%%%%%%%%%%%%%%%%%%%%%%%%%%%%%%%%%%%%%%%%%%%%%%%%%%%%%%%%%%%%%%%%%%%%%%%%%%%%%%%%

\begin{figure}[ht]
\centering
\includegraphics[width=0.99\columnwidth]{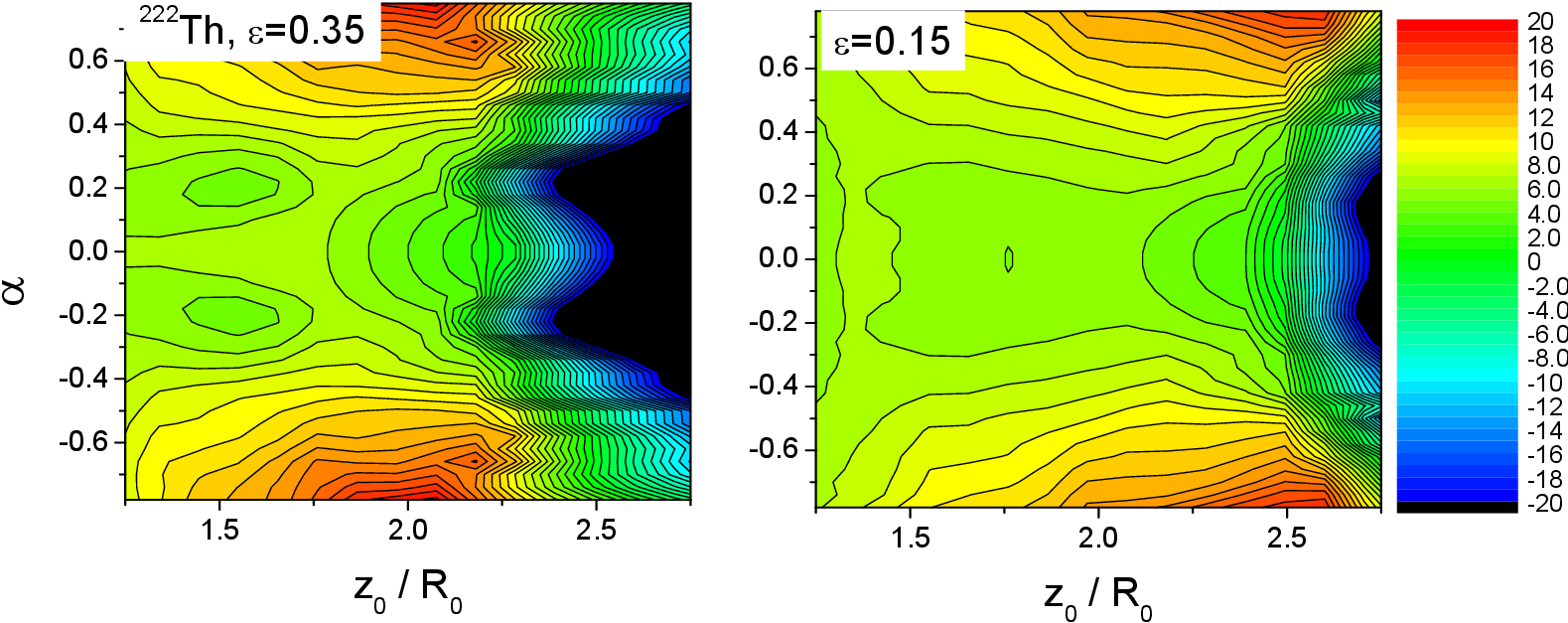}
\caption{The dependence of the potential energy of $^{222}$Th, minimized with respect to $\delta_1, \delta_2$ at fixed $\eps$, $\eps$=0.35 and $\eps$=0.15, on the elongation $z_0/R_0$ and asymmetry $\alpha$.}
\label{edef222min}
\end{figure}

The comparison of the total deformation energy of $^{222}$Th at the saddle and above in the ($z_0, \alpha$) plane is shown in Fig.\,\ref{edef222min}. In this plot the energy was minimized with respect to $\delta_1, \delta_2$ at fixed $\eps$ for $\eps$=0.35 and $\eps$=0.15. For $\eps$=0.35 one clearly sees the two asymmetric valleys leading to asymmetric fission. For $\eps$=0.15 the mass asymmetric valley is less pronounced. So symmetric divisions may also occur. These observations are in accord with the fission fragment mass distribution shown in the left hand part of Fig.\,\ref{4D_Chatil}.
%%%%%%%%%%%%%%%%%%%%%%%%%%%%%%%%%%%%%%%%%%%%%%%%%%%%%%%%%%%%%%%%%
%%%%%%%%%%%%%%%%%%%%%%%%%%%%%%%%%%%%%%%%%%%%%%%%%%%%%%%%%%%%%%%%%%%%%%%%%%%%%%%%%%%%%
%%%%%%%%%%%%%%%%%%%%%%%%%%%%%%%%%%%%%%%%%%%%%%%%%%%%%%%%%%%%%%%%%%%%%%%%%%%%%%%%%%%%%%%%%%%%%%%%%%%%%%%%%%%%%%%%%%%%%%%%%%%%%%%%%%%%%%%%%%%%%%%%%%%%%%%%%%%%%%%%%%%%%%%%%%%%%%%%%%%%%%%%%%%%%%%%%%%%%%%%%%%%%%
\section{The tensors of friction and inertia}
\label{transp}
%%%%%%%%%%%%%%%%%%%%%%%%%%%%%%%%%%%%%%%%%%%%%%%%%%%%%%%%%%%%%%%%%
In the present work we use the macroscopic transport coefficients which are often used for solving the Langevin equations. The macroscopic transport coefficients depend only on the shape of the system and do not depend on the excitation energy of the system.

The macroscopic mass tensor $M_{\mu\nu}^{WW}$ is usually defined in Werner-Wheeler approximation \cite{werwhe},
\begin{equation}\label{massww}
M_{\mu\nu}^{WW} =
 \pi \rho_{_0} \int \rho^2(z) \left[ A_{\mu}(z) A_{\nu}(z)
+ \frac{\rho^2}{8} A_{\mu}'(z) A_{\nu}'(z) \right] dz,
\end{equation}
where $\rho(z)$ is the profile function of the axially symmetric shape and
\begin{equation}\label{amyu}
A_{\mu}(z;Q)=\frac{1}{\rho^2(z,Q)}\frac{\partial}{\partial Q_{\mu}} \int_{z}^{z_{_R}} \rho^2 (z',Q) dz'-\frac{\partial z_{cm}}{\partial Q_{\mu}}.
\end{equation}
The derivative of center-of-mass $\partial z_{cm}/\partial Q_{\mu}$ is subtracted in (\ref{amyu}) from the standard definition of $A_{\mu}(z;Q)$ to exclude from the velocity field the spurious contribution due to the center-of-mass motion.

The macroscopic friction tensor is given by the so-called the wall-and-window formula \cite{runswiat}.
The wall-and-window friction is a generalization of wall friction \cite{wall1} which for
axial symmetric shapes can be written as \cite{krapom},
%\begin{equation}\label{frwall}
%\gamma_{\mu\nu}^{\mathrm{wall}}=\pi\rho_{_0} v_F\int_{z_{_L}}^{z_{_R}}dz \frac{\partial \rho^2/\partial Q_{\mu}\,\,\partial %\rho^2/\partial Q_{\nu}}{\sqrt{4\rho^2+(\partial \rho^2 / \partial z)^2}}\,,
%\end{equation}
\begin{equation}\label{frwall}
\gamma_{\mu\nu}^{\mathrm{wall}}=\pi\rho_{_0} \bar v\int dz \frac{\left(\frac{\partial \rho^2}{\partial Q_{\mu}}+\frac{\partial \rho^2}{\partial z}\frac{\partial z_{cm}}{\partial Q_{\mu}}\right)\left(\frac{\partial \rho^2}{\partial Q_{\nu}}+\frac{\partial \rho^2}{\partial z}\frac{\partial z_{cm}}{\partial Q_{\nu}}\right)}{\sqrt{4\rho^2+(\partial \rho^2 / \partial z)^2}}\,,
\end{equation}
with $\rho_0=3mA/(4\pi R_0^3)$ and $\bar v$ being the mean velocity of nucleons. For $\bar v$ the Fermi gas estimate was used, $\bar v=(3/4)v_F$, where $v_F$ is the Fermi velocity, related to the particle number $A$ by Thomas-Fermi relation $A=(4/9\pi) x_F^3, x_F\equiv m v_F/\hbar$. The derivatives $\partial \rho^2/\partial Q$ in Eq.\,(5.172) of \cite{krapom} should be calculated under the condition that the nuclear center of mass remains fixed during the fission process.
 This requirement is fulfilled by adding terms, proportional to $\partial z_{cm}/\partial Q$, in the numerator of (\ref{frwall}).

In the wall-and-window model the velocities of nucleons are taken with respect to the velocities of centers of mass of the left or right parts of nucleus and a "window"  term was added \cite{runswiat,krapom},
\begin{equation}\label{frwall2}
\gamma_{\mu\nu}^{\mathrm{w+w}}=\pi\rho_{_0}\bar v\left(\int_{z_{L}}^{0}I_L(z)\,dz+\int_{0}^{z_{_R}}I_R(z)dz \right)
+\gamma_{\mu\nu}^{\mathrm{window}},
\end{equation}
with
\begin{equation}\label{ILR}
I_{L,R}(z)=\frac{\left(\frac{\partial \rho^2}{\partial Q_{\mu}}+\frac{\partial \rho^2}{\partial z}\frac{\partial z_{cm}(L,R)}{\partial Q_{\mu}}\right)\left(\frac{\partial \rho^2}{\partial Q_{\nu}}+\frac{\partial \rho^2}{\partial z}\frac{\partial z_{cm}(L,R)}{\partial Q_{\nu}}\right)
}{\sqrt{4\rho^2+(\partial \rho^2 / \partial z)^2}},
\end{equation}
and
\begin{equation}\label{frwindow}
\gamma_{\mu\nu}^{\mathrm{window}}=
\frac{1}{2}\rho_0\bar v\left[\Delta\sigma\left(\frac{\partial R_{12}}{\partial q_{\mu}}\frac{\partial R_{12}}{\partial q_{\nu}}\right)+\frac{32}{9\Delta\sigma}\frac{\partial V_L}{\partial q_{\mu}}\frac{\partial V_L}{\partial q_{\nu}}\right] \ ,
\end{equation}
where $R_{12}$ is the distance between the centers of mass of the left and right parts of nucleus and $\Delta\sigma$ is the area of the "window".

One expects a smooth transition between the regime in which the wall formula applies and the part of fission path where wall-and-window friction should be used. For this Nix and Sierk \cite{nixsierk} proposed the phenomenological ansatz
\begin{equation}\label{total}
\gamma_{\mu\nu}^{total}=\sin^2(\pi\phi/2)\gamma_{\mu\nu}^{wall}+\cos^2(\pi\phi/2)\gamma_{\mu\nu}^{w+w},
\end{equation}
with $\phi=(r_{neck}/R_{min})^2$, where $R_{min}$ is the minimal semi-axes of the two outer ellipsoidal shapes.

Soon after the introduction of wall friction it was recognised that the wall friction is too strong and the reduction factor, $k_s=0.27$, was introduced by Nix and Sierk \cite{nix1987} from the analysis of widths of giant resonances. This reduction factor was used in all our Langevin calculations.

\begin{figure}[ht]
\centering
\includegraphics[width=0.99\columnwidth]{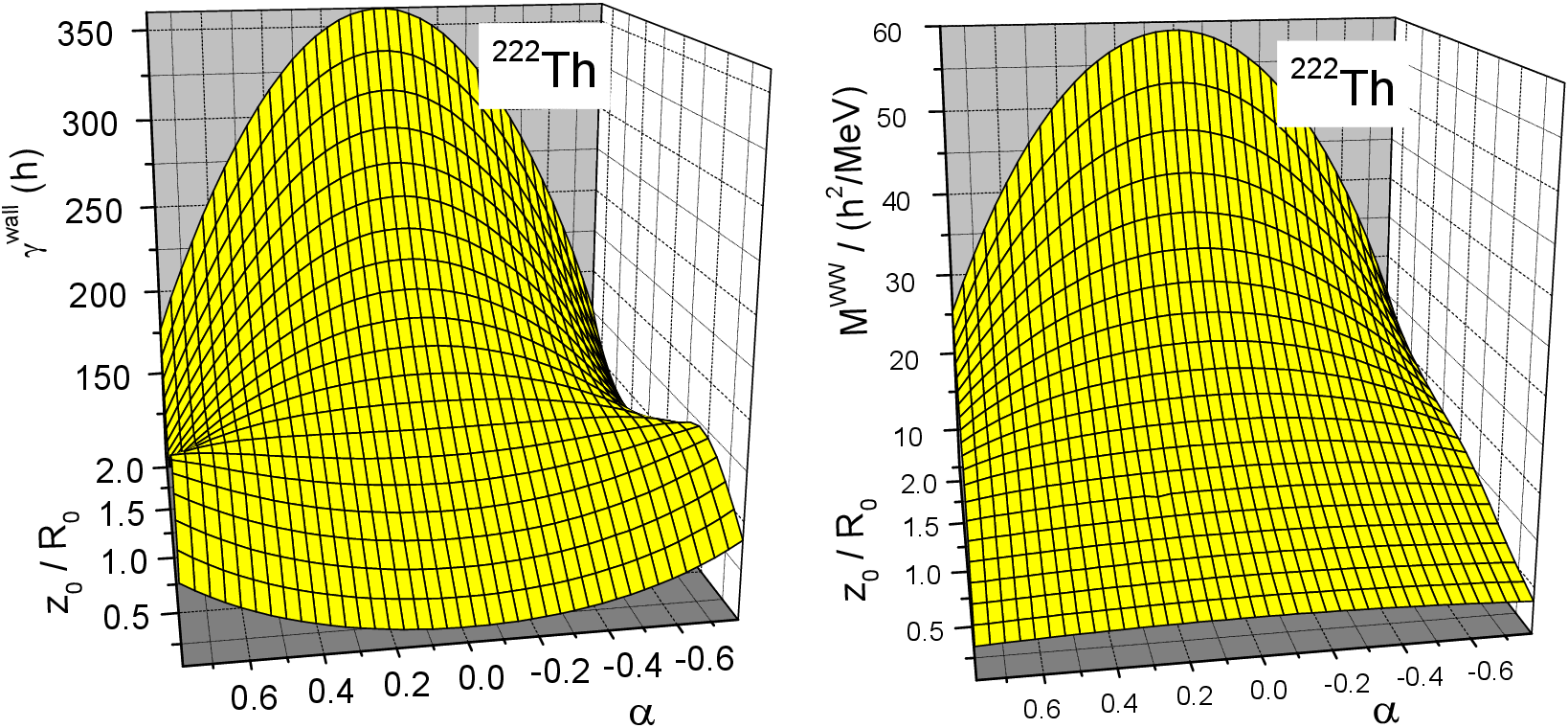}
\caption{The $z_0z_0$ components of friction (\ref{frwall}) and inertia (\ref{massww}) tensors for $^{222}$Th as function of the elongation $z_0/R_0$ and mass asymmetry $\alpha$ at fixed $\delta_1=\delta_2$=0.2 and $\eps$=0.35.}
\label{transp222}
\end{figure}

The $z_0z_0$-components of the friction (\ref{frwall}) and inertia (\ref{massww}) tensors for $^{222}$Th are shown in Fig.\,\ref{transp222} as function of the elongation $z_0/R_0$ and mass asymmetry $\alpha$. The dependence of friction or inertia separately on the deformation does not have much meaning. This dependence is a consequence of the definition of the deformation parameters.
More meaningful is the ratio of friction to inertia, the so-called reduced friction coefficient, $\beta_{\mu\nu}\equiv(\gamma M^{-1})_{\mu\nu}$,
or the damping parameter $\eta_{\mu\nu}\equiv\beta_{\mu\nu}/2\varpi$, where $\varpi$ is the frequency of local collective vibrations. This is a dimensionless parameter that defines whether the collective motion is under-damped, $\eta <<$1, or over-damped $\eta >>$1. Accordingly, one develops different models for the under-damped and over-damped motions.
In the model by Kramers \cite{kramers} one can find the expressions for the limits of
high viscosity,$\Gamma_{HV}$, and low viscosity, $\Gamma_{LV}$.
 These expressions are rather different.
%\bel{GammaK}
%\Gamma_{HV}=\frac{\hbar\varpi}{2\pi}e^{-V_b/T}(\sqrt{1+\eta^2}-\eta)\,,\quad
%\Gamma_{LV}=\frac{\hbar\bar\gamma}{M}\frac{V_b}{T}e^{-V_b/T}.
%\end{equation}
%The high viscosity limit corresponds to large $\eta$, $\eta>> 1$. In the low viscosity limit $\eta<< 1$.
So, it is important to know whether the nuclear fission process is under-damped or over-damped in order to develop the meaningful models.

\begin{figure}[ht]
\centering
\includegraphics[width=0.99\columnwidth]{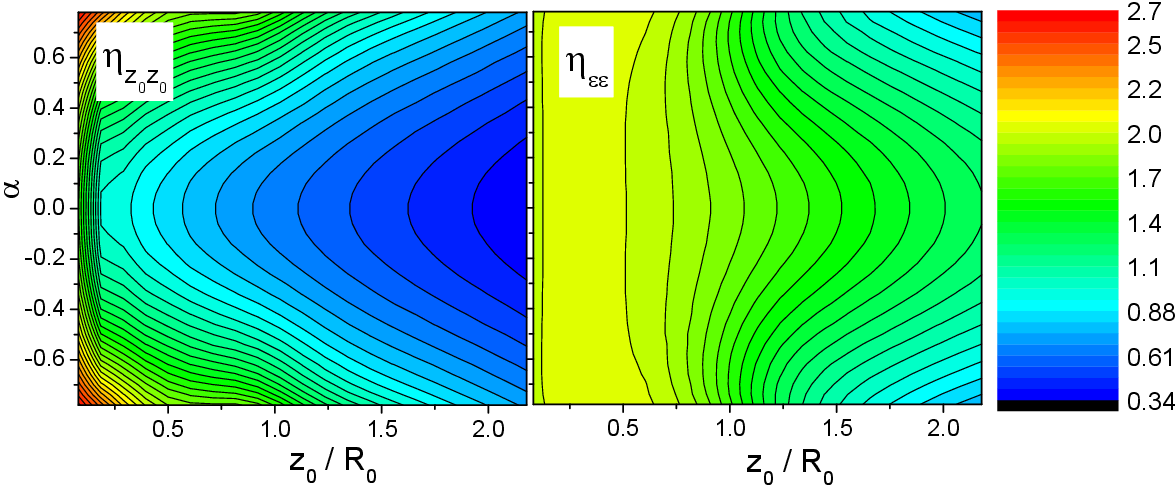}
\caption{The damping parameters $\eta_{z_0z_0}$ and $\eta_{\eps\eps}$ for $^{222}$Th as function of the elongation $z_0/R_0$ and mass asymmetry $\alpha$ at fixed $\delta_1=\delta_2$=0.2 and $\eps$=0.35.}
\label{beta}
\end{figure}

In our calculation we use the reduced value of wall and window friction with the reduction factor $k_s$=0.27. The damping factor shown in Fig.\,\ref{beta} includes this factor, $\eta_{\mu\nu}=k_s(\gamma M^{-1})_{\mu\nu}/2\varpi$.
%%%%%%%%%%%%%%%%%%%%%%%%%%%%%%%%%%%%%%%%%%%%%%%%%%%%%%%%%%%%%%%%%%%%%%%%%%%%%%%%%%%%%
%%%%%%%%%%%%%%%%%%%%%%%%%%%%%%%%%%%%%%%%%%%%%%%%%%%%%%%%%%%%%%%%%%%%%%%%%%%%%%%%%%%%%

The $z_0z_0$ and $\eps\eps$ components of damping tensor $\eta_{\mu\nu}$ for $^{222}$Th are shown in Fig.\,\ref{beta} for the deformations at the saddle and above. As one can see, $\eta_{\mu\nu}$ is more stable with respect to deformation as compared to $\gamma$ or $M$ alone. The value of $\eta_{z_0z_0}$ varies in the limits from 2.6 to 0.35. The value of $\eta_{\eps\eps}$ varies in the limits from 1.9 to 1.3. Thus, in the present Langevin treatment with macroscopic transport coefficient the fission process is neither under-damped nor over-damped. So, the application of the Kramers' high viscosity limit
%$\Gamma_{HV}$ (\ref{GammaK}) of
formula to the nuclear fission width may not be justified.

This conclusion is correct under the assumption that the underlying approximations are reliable. For the damping parameter shown in Fig.\,\ref{beta} we used the Werner-Wheeler approximation for the inertia, the reduced wall friction with the reduction factor $k_s$=0.27, and the frequency $\varpi$=2 MeV/$\hbar$. For other parameter values the damping factor may be different.
%%%%%%%%%%%%%%%%%%%%%%%%%%%%%%%%%%%%%%%%%%%%%%%%%%%%%%%%%%%%%%%%%
%%%%%%%%%%%%%%%%%%%%%%%%%%%%%%%%%%%%%%%%%%%%%%%%%%%%%%%%%%%%%%%%%%%%%%%%%%%%%%%%%%%%%

%%%%%%%%%%%%%%%%%%%%%%%%%%%%%%%%%%%%%%%%%%%%%%%%%%%%%%%%%%%%%%%%%%%%%%%%%%%%%%%%%%%%%%%%%%%%%%%%%%%%%%%
\section{Numerical results}
\label{results}
%%%%%%%%%%%%%%%%%%%%%%%%%%%%%%%%%%%%%%%%%%%%%%%%%%%%%%%%%%%%%%%%%
In this Section we present the results of the numerical solution of the five-dimensional Langevin equations (\ref{lange}) with the potential energy and transport coefficients specified above. In these calculations the damping of shell correction with the excitation energy was described by the method developed in \cite{shcot}. The frequency of local collective vibration was assumed to be constant, $\hbar\varpi$=2 MeV. In principle, $\varpi$ should depend on the deformation. But, the account of this dependence makes the computation time too time consuming.

\begin{figure}[ht]
\centering
\includegraphics[width=0.95\columnwidth]{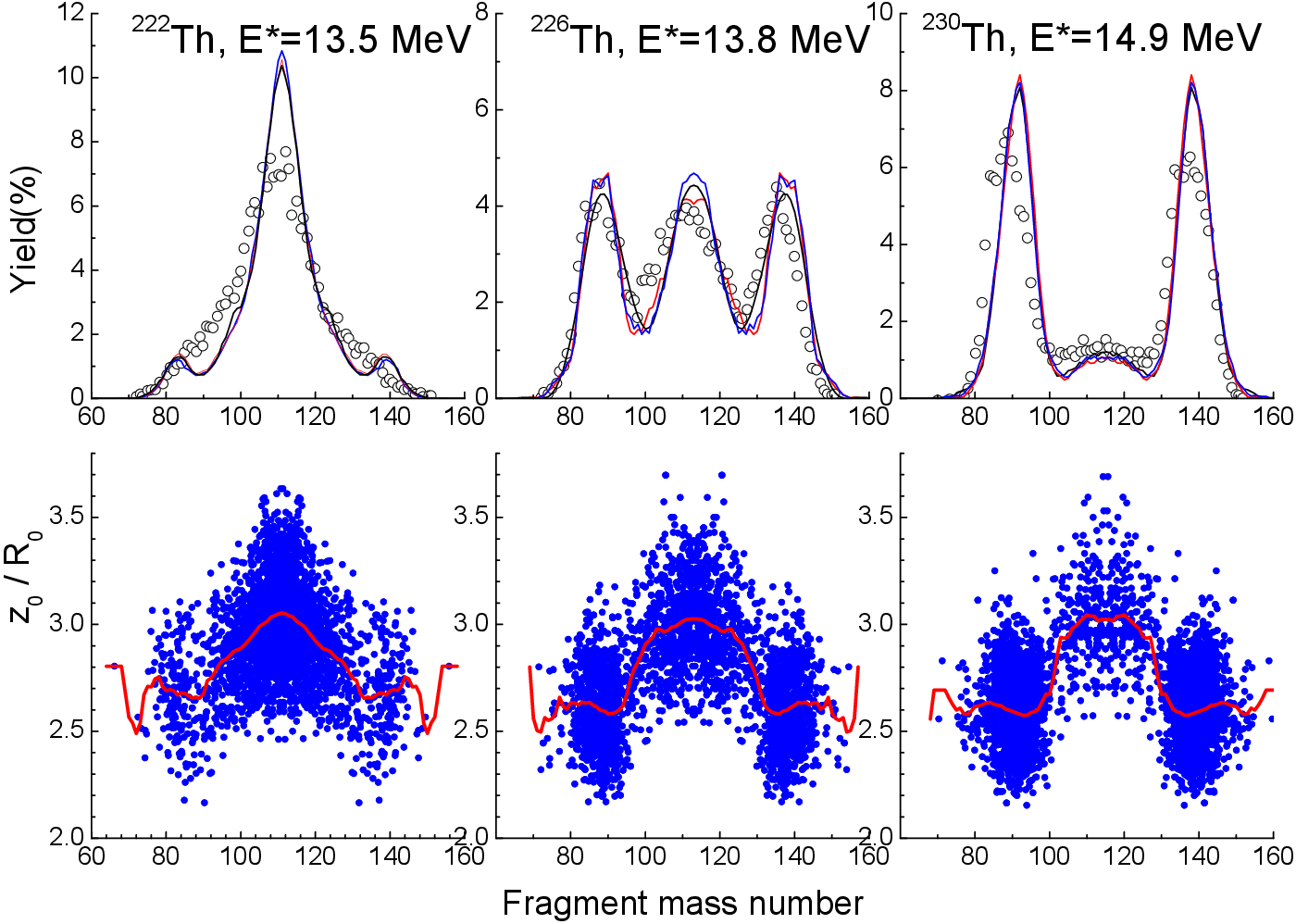}
\caption{Top: The comparison of calculated fission fragments mass distributions (solid lines) for a few thorium isotopes with the experimental data \protect\cite{chatil1} (open circles). The mass distributions calculated with $\hbar\varpi=0 \,\,(T^*=T)$, $\hbar\varpi$=2 MeV and  $\hbar\varpi$=4 MeV are shown by red, black and blue lines correspondingly.
Bottom: The distribution of fission events in the elongation-asymmetry plane.}
\label{5D_Chatil}
\end{figure}

To justify the choice of $\hbar \varpi$, we have carried out the calculations with
three values of $\hbar \varpi$: $\varpi$=0 ($T^{\ast}=T$), $\hbar \varpi$=2 and $\hbar \varpi$=4
Mev, see the red, black and blue curves in Fig.\,\ref{5D_Chatil}. The mass distributions
calculated with all three values of $\varpi$ are almost identical. Thus, the choice $\hbar \varpi$=2 MeV is quite reliable.

For solving of Eqs.~(1-2), besides the coefficients of equations, one should fix the initial values and final conditions. The calculations for all thorium isotopes were started from the same point close to the ground state deformation, $\{q_{\mu}\}$={0.2, 0.2, 0.2, 0.0, 0.35} with zero collective momenta $\{p_{\mu}\}=0$. The integration of equations continued until the neck radius turned into $r_{neck}^{(crit)}$=1 fm. At this point the solutions of the equations provide all the information on the shape, collective velocities and excitation energy of the system. This information makes it possible to calculate the moments of the density distribution, the mass distribution of fission fragments, the prescission and total kinetic energies, excitation energy of nucleus just before the scission.
%%%%%%%%%%%%%%%%%%%%%%%%%%%%%%%%%%%%%%%%%%%%%%%%%%%%%%%%%%%%%%%%%
%%%%%%%%%%%%%%%%%%%%%%%%%%%%%%%%%%%%%%%%%%%%%%%%%%%%%%%%%%%%%%%%%%%%%%%%%%%%%%%%%%%%%

Unfortunately, the 5D computations are too time consuming. The 5D configuration space is much larger compared with 4D space. The trajectories travel in 5D space much longer before they reach the scission point.

Another problem is the lack of the memory space. For the grid in $z_0, \delta_1, \delta_2, \alpha$ that we use usually, $41\times 31\times 31\times 40$ we can add only 5 points in $\eps$. Otherwise the memory of computer is overfilled. For more points in $\eps$ we have to reduce the grid in other deformation parameters. In present calculations we used, $0.1\leq\eps\leq 0.5$ with $\Delta\eps$=0.1.

The distribution of fission events in the elongation-asym\-met\-ry plane is shown in the bottom part of Fig.\,\ref{5D_Chatil}. The red line is the average value of $z_0/R_0$. One can see that symmetric fission occurs at much more elongated configuration than mass-asymmetric. The symmetric fission mode could
therefore be called super-long. The elongation of this mode is almost the same for all three isotopes. Thus, our calculations do not confirm the  conclusion made in \cite{chatil2} from the analysis of multiplicity of post-scission neutrons, that the symmetric fission mode in $^{222}$Th is super-short.

For a more definite conclusion it is necessary to carry out 5D calculations of more observables of the fission process, the charge and kinetic energy yields, fragment mean N/Z, multiple moments and the excitation energies of fragments, postscission neutron multiplicities and their correlations.
This will help to investigate the interpretation of the measurement since the additional observables will be
a better probe of the deformation at scission. Such calculations will be the subject of our future work.
Recently, similar 4D calculations with Fourier-over-spheroid shape parametrization were published by Pomorski ${\it at~ al} $ \cite{pomo2023}.
%%%%%%%%%%%%%%%%%%%%%%%%%%%%%%%%%%%%%%%%%%%%%%%%%%%%%%%%%%%%%%%%%
\begin{figure}[ht]
\centering
\includegraphics[width=0.95\columnwidth]{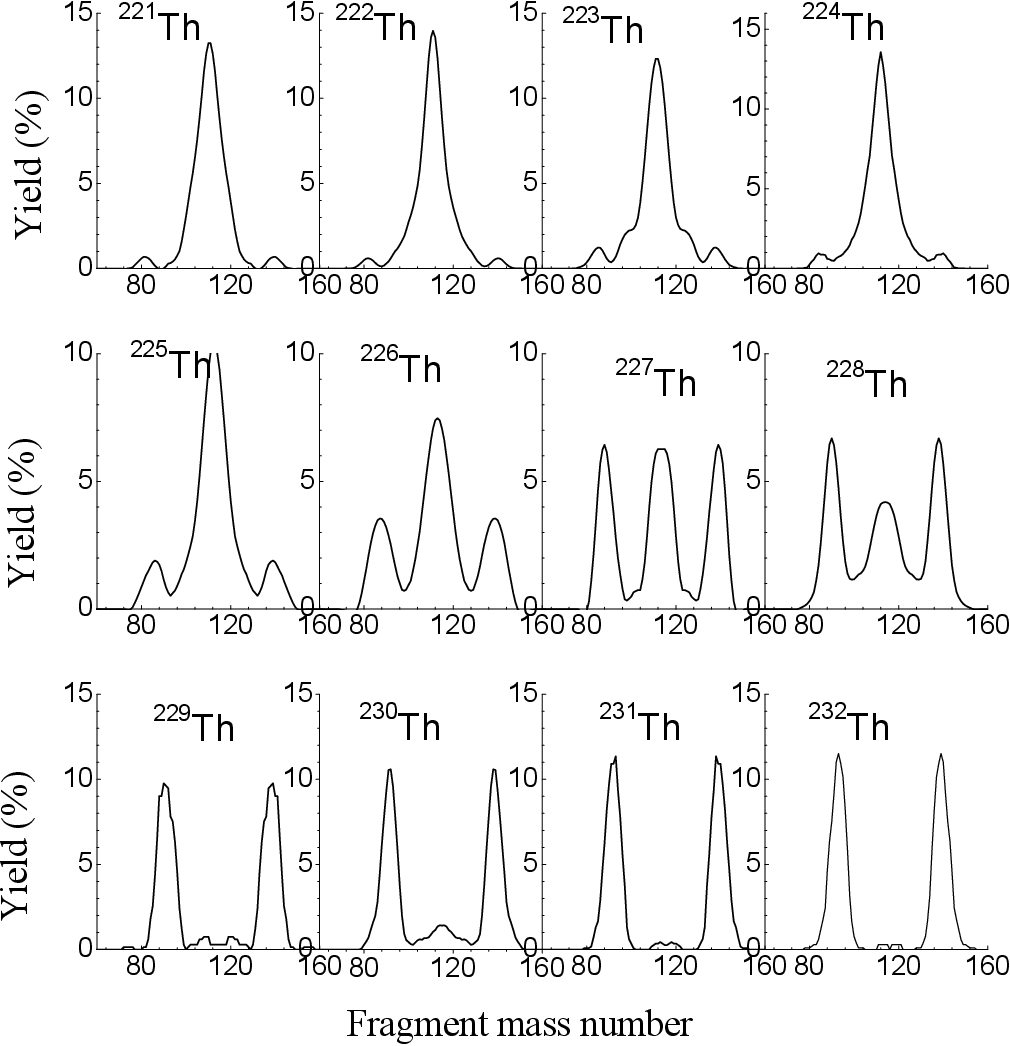}
\caption{The calculated mass distributions of fission fragments for the sequence of thorium isotopes at the excitation energy $E^*$=11 MeV.}
\label{Th_chain11}
\end{figure}
%%%%%%%%%%%%%%%%%%%%%%%%%%%%%%%%%%%%%%%%%%%%%%%%%%%%%%%%%%%%%%%%%%%%%%%%%%%%%%%%%%%%%

The calculated mass distributions of fission fragments for the sequence of thorium isotopes at the excitation energy $E^*$=11 MeV are shown in Fig.\,\ref{Th_chain11}. The calculated distributions agree qualitatively with the experimental results of K.-H. Schmidt ${\it at~ al} $ \cite{khs}.

%%%%%%%%%%%%%%%%%%%%%%%%%%%%%%%%%%%%%%%%%%%%%%%%%%%%%%%%%%%%%%%%%%%%%%%%%%%%%%%%%%
\section{Summary}
%%%%%%%%%%%%%%%%%%%%%%%%%%%%%%%%%%%%%%%%%%%%%%%%%%%%%%%%%%%%%%%%%%%%%%%%%%%%%%%%%%
%{\bf Summary.}
The calculations within the 5-dimensional dynamical Langevin approach provide much better agreement with the available experimental data as compared with 4D calculations. In particular, the transition from the mass-symmetric to mass-asymmetric fission via the triple-humped distribution in fission of thorium isotopes is well reproduced.

%It was shown that the mass-symmetric mode in fission of thorium isotopes is superlong and mass-symmetric mode standard. These modes are clearly seen in the distributions of total kinetic energies of fission fragments. The calculated results do not confirm that the symmetric scission in light thorium isotopes shows a compact configuration.

The use of 5-dimensional Langevin calculations makes theoretical predictions for the observables of fission process much more reliable.

%\begin{acknowledgments}
{\bf Acknowledgments.}
The authors would like to express their gratitude to Profs. K. Pomorski, J. Randrup, and C. Schmitt for valuable comments and suggestions.
%\end{acknowledgments}

\end{document}